\documentclass[12pt, draftcls, onecolumn]{IEEEtran}
\usepackage{amsmath}
\usepackage{amssymb}
\usepackage{amsfonts}
\usepackage{amscd}
\usepackage{mathrsfs}
\usepackage{graphicx}
\usepackage{color}
\usepackage{url}
\usepackage{bm}
\usepackage{algorithm}
\usepackage{algorithmic}

\newtheorem{example}{Example}
\newtheorem{remark}{Remark}

\begin{document}

\title{Integer Space-Time Block Codes for\\ Practical MIMO Systems}
\author{\authorblockN{J.~Harshan and E.~Viterbo}\\\thanks{This work was performed at the Monash Software Defined Telecommunications (SDT) Lab and was supported by the Monash Professional Fellowship and the Australian Research Council under Discovery grants ARC DP 130100103.} 
\authorblockA{Department of Electrical and Computer Systems Engineering,\\
Monash University, Victoria, Australia}\\
{\small {\tt $\{$harshan.jagadeesh, emanuele.viterbo$\}$@monash.edu}}}
\and

\maketitle

\begin{abstract}
Full-rate space-time block codes (STBCs) achieve high spectral-efficiency by transmitting linear combinations of information symbols through every transmit antenna. However, the coefficients used for the linear combinations, if not chosen carefully, results in ({\em i}) large number of processor bits for the encoder and ({\em ii}) high peak-to-average power ratio (PAPR) values. In this work, we propose a new class of full-rate STBCs called Integer STBCs (ICs) for multiple-input multiple-output (MIMO) fading channels. A unique property of ICs is the presence of integer coefficients in the code structure which enables reduced numbers of processor bits for the encoder and lower PAPR values. We show that the reduction in the number of processor bits is significant for small MIMO channels, while the reduction in the PAPR is significant for large MIMO channels. We also highlight the advantages of the proposed codes in comparison with the well known full-rate algebraic STBCs.
\end{abstract}

\section{Introduction and Preliminaries}
\label{sec1}

Space-time coding is a powerful technique for communication over multiple-input multiple-output (MIMO) fading channels. It is effective not only to combat the degrading effects of multi-path fading \cite{TSC} but also to provide increased data rate \cite{Tel}. For an $n \times n$ MIMO channel (with $n$ transmit antennas and $n$ receive antennas), a space-time block code (STBC), if appropriately designed is known to provide a rate of $n$ complex symbols per channel use \cite{Tel}, and a spatial diversity of $n^2$ \cite{TSC}. Such STBCs are referred to as full-rate, full-diversity STBCs. 

In this paper, we address the implementation aspects of full-rate, full-diversity STBCs with emphasis on the following two metrics: ({\em i}) the number of bits for quantization \cite{Walden} at the transmitter, which impacts the overall memory size, and ({\em ii}) the PAPR, which influences the total power consumption \cite{brian}. 
We now explain the importance of incorporating quantization effects in the code design, and then address the PAPR issues. Consider the well known rate-$1$ Alamouti design \cite{YVXC} given by,
\begin{equation}
\label{alamouti}
\textbf{X}_{A} = \left[\begin{array}{rr}
x_{1} & -x_{2}^{*}\\
x_{2}  & x_{1}^{*}\\
\end{array}\right],
\end{equation}
where the variables $x_{1}, x_{2}$ take values from an underlying complex constellation $\mathcal{S}$ (e.g. QAM). For this case, encoding an STBC codeword is straightforward, as the input bits (either uncoded or coded bits) are mapped to the symbols of $\mathcal{S}$, and the corresponding symbols are transmitted in the format of \eqref{alamouti}. Now, consider the rate-$2$ Golden code \cite{BRV} design given by,
\begin{equation}
\label{gcode}
\textbf{X}_{G} = \left[\begin{array}{rr}
f_{1}(x_{1}, x_{2}) & f_{2}(x_{3}, x_{4})\\
f_{3}(x_{3}, x_{4})  & f_{4}(x_{1}, x_{2})\\
\end{array}\right],
\end{equation}
where $x_{1}, x_{2}, x_{3}, x_{4} \in \mathcal{S}$ carry information symbols, and $\{ f_{j}(\cdot)\}$ are some \emph{irrational} linear combinations of these symbols \cite{BRV}. For every codeword transmission, the encoding operation for the Golden code involves computation of linear combinations of information symbols with irrational coefficients. 
An alternate technique is to store all the codewords of the code, and use the repository as a look up table (LUT) to map the information bits to the codewords. However, this method has exponential memory requirements as the constellation size increases. When the Golden code is implemented with limited hardware, it has been shown in \cite{HaV1} that there will be substantial degradation in the error-performance due to the finite-precision encoding operations. Apart from the Golden code, all other well known full-rate algebraic STBCs \cite{DTB}-\cite{PaS3} have irrational coefficients in the code structure, which leads to degradation in the performance (with limited hardware).\\ 
\indent In addition to the above problem, transmitting linear combination of information symbols also increases the PAPR values if the coefficients are not carefully chosen. For linear power amplifiers, a high dynamic range in the transmitter hardware is expensive and consumes more power. On the other hand, non-linear amplifiers results in spectral spread leading to out of band radiation and in-band distortion \cite{brian}. Importantly, the PAPR disadvantages are severe for large MIMO channels as the number of information symbols to be linearly combined increases linearly with the number of transmit antennas.\\ 
\indent The above mentioned problems have motivated us to design a new class of full-rate STBCs with few bits for encoding operations, and low PAPR values. For some prior works on the effects of quantization at the receiver for fading channels, we refer the reader to \cite{YiL}-\cite{YLC} and the references within. Henceforth, we only consider transmitter side quantization effects and assume that the receiver works with infinite precision. This work stems from the preliminary observations made in \cite{HaV1} and generalizes the code construction in \cite{HaV2} to any $n \times n$ MIMO channel.

\section{Signal Model}
\label{sec2}

The $n \times n$ MIMO channel, denoted by the matrix $\textbf{H}$, consists of a source and a destination terminal each equipped with $n$ antennas. For $i, j\in \{1, 2, \ldots, n\}$, the channel between the $i$-th transmit antenna and the $j$-th receive antenna is assumed to be flat fading and hence, denoted by $\textbf{H}_{i, j}$. Each $\textbf{H}_{i, j}$ remains constant for a block of $T$ ($T \geq n$) complex channel uses and is assumed to take an independent realization in the next block. Statistically, we assume $\textbf{H}_{i, j} \sim ~\mathcal{N}_{c}(0, 1) ~\forall i, j$ across quasi-static intervals. The source conveys information to the destination through an $n \times n$ full-rate STBC denoted by $\mathcal{C}$. We assume that a linear design $\textbf{X}_{\mathcal{LD}}$ in complex variables $x_{r,s}$ for $r, s\in \{1, 2, \ldots, n\}$ (a total of $n^{2}$ symbols) is used to generate $\mathcal{C}$ by taking values from an underlying complex constellation $\mathcal{S}$. If $\textbf{X} \in \mathcal{C}$ denotes a transmitted codeword matrix such that $E[|\textbf{X}_{i, j}|^{2}]= P_{s} ~\forall i, j$, then the received matrix $\textbf{Y} \in \mathbb{C}^{n \times n}$ at the destination is given by,
\begin{equation*}
\label{MIMO_channel}
\textbf{Y} = \sqrt{\frac{1}{n}}\textbf{H}\textbf{X} + \textbf{Z},
\end{equation*}
where $\textbf{H} \in \mathbb{C}^{n \times n}$ denotes the channel matrix, $\textbf{Z} \in \mathbb{C}^{n \times n}$ denotes the AWGN with its entries distributed as $\mathcal{N}_{c}(0, \sigma^2)$. With this, the average receive signal power-to-noise ratio (SNR) per receive antenna is $P_{s}/\sigma^2$. If $\eta = \max |\textbf{X}_{i, j}|^{2}/E[|\textbf{X}_{i, j}|^{2}]$ denotes the PAPR, then we define the peak signal power-to-noise ratio (denoted by PSNR) as $\eta P_{s}/\sigma^2$. We assume a coherent MIMO channel where only the receiver has the complete knowledge of $\textbf{H}$. In the rest of this section, we discuss the structure of fixed-point encoding operations for the STBCs.

\indent We represent the parameters of the design $\textbf{X}_{\mathcal{LD}}$ and the symbols of $\mathcal{S}$ using $(q-1)$-bits for the fractional part and $1$-bit for the sign part. For any $y \in \mathbb{R}$ such that $|y| \leq 1$, we define the $q$-bit quantized version of $y$ as $Q_{q}(y) \triangleq \lfloor y2^{q-1} \rceil/2^{q-1},$ where $\lfloor x \rceil$ represents the nearest integer to $x$. To avoid quantization errors due to \emph{over-flows}, we scale down all the parameters using a suitable scale factor such that each component falls within the quantization interval $[-1, 1)$. With the above fixed-point representation, the $q$-bit encoding operation is only subject to quantization errors due to rounding operations in the least significant bits.

\section{Integer STBCs}
\label{sec4}
We now propose full-rate STBCs which need reduced number of bits for encoding operations with $M$-QAM. These codes do not contain irrational coefficients, and hence, can be exactly represented in finite-precision. We refer to such codes as Integer STBCs (ICs) and they can be obtained by the linear design 

\begin{small}
\begin{equation}
\label{integer_code_design}
\textbf{X}_{I} = \left[\begin{array}{ccccccc}
<\Phi_{1}, \textbf{x}_{1}> & <\Phi_{1}, \textbf{x}_{2}> & <\Phi_{1}, \textbf{x}_{3}> & \cdots & <\Phi_{1}, \textbf{x}_{n-1}> & <\Phi_{1}, \textbf{x}_{n}>\\
\gamma <\Phi_{2}, \textbf{x}_{n}> & <\Phi_{2}, \textbf{x}_{1}> & <\Phi_{2}, \textbf{x}_{2}> & \cdots & <\Phi_{2}, \textbf{x}_{n-2}> & <\Phi_{2}, \textbf{x}_{n-1}>\\
\vdots & & & \vdots & \vdots & \vdots &\\
\gamma <\Phi_{n-1}, \textbf{x}_{3}> & \gamma <\Phi_{n-1}, \textbf{x}_{4}> & \gamma <\Phi_{n-1}, \textbf{x}_{5}> & \cdots & <\Phi_{n-1}, \textbf{x}_{1}> & <\Phi_{n-1}, \textbf{x}_{2}>\\
\gamma <\Phi_{n}, \textbf{x}_{2}> & \gamma <\Phi_{n}, \textbf{x}_{3}> & \gamma <\Phi_{n}, \textbf{x}_{4}> & \cdots & \gamma <\Phi_{n}, \textbf{x}_{n}> & <\Phi_{n}, \textbf{x}_{1}>\\
\end{array}\right],
\end{equation}
\end{small}
%

\noindent where for $j, k \in \{1, 2, \ldots, n \}$, $<\Phi_{k}, \textbf{x}_{j}>$ denotes the inner product of the $k$-th row of the circulant matrix 
\begin{equation*}
\Phi \triangleq \left[\begin{array}{cccccc}
1 & \alpha & \alpha^2 & \cdots & \alpha^{n-2} & \alpha^{n-1}\\
\alpha^{n-1} & 1 & \alpha & \cdots & \alpha^{n-3} & \alpha^{n-2}\\
\vdots & \vdots &  & \vdots & \vdots & \vdots\\
\alpha^2 & \alpha^3 & \alpha^4 & \cdots & 1 & \alpha\\
\alpha & \alpha^2 & \alpha^3 & \cdots & \alpha^{n-1} & 1\\
\end{array}\right],
\end{equation*}
and the vector $\textbf{x}_{j} = [x_{j,1}~ x_{j,2}~ x_{j,3}~ \ldots~ x_{j,n}].$ Here, $\gamma = \imath$ $\left( \mbox{where } \imath = \sqrt{-1} \right)$ and $\alpha$ is an appropriately chosen integer. The variables $x_{j, 1}, x_{j, 2}, \ldots, x_{j, n}$ carry information symbols from a square $M$-QAM given by

\begin{small}
\begin{equation*}
\mathcal{S} = \left\lbrace a + b \imath ~|~ a, b \in \{ -2^{\frac{m}{2}} + 1, \ldots, -1, 1, \ldots, 2^{\frac{m}{2}} - 1\} \right\rbrace,
\end{equation*}
\end{small}

\noindent where $M = 2^{m}$ and $m$ is even. For the $2^{m}$-QAM constellation, we choose $\alpha = 2^{\frac{m}{2}}$ in the Integer STBCs. Note that we do not use phase shift keying (PSK) constellations as the underlying complex symbols require larger number of bits for quantization. 

\begin{example}
For $2 \times 2$ MIMO, an Integer STBC can be obtained using the design,
\begin{equation}
\label{integer_code_design_2_X_2}
\left[\begin{array}{cc}
x_{1,1} + \alpha x_{1,2} & x_{2,1} + \alpha x_{2,2}\\
\gamma (\alpha x_{2,1} + x_{2,2}) & \alpha x_{1,1} + x_{1,2}\\
\end{array}\right].
\end{equation}
For the above design, the parameter $\alpha$ takes the value of $2$, $4$, and $8$ when the variables $x_{1,1}, x_{1,2}, \ldots, x_{2,2}$ take values from $4$, $16$, and $64$-QAM constellations, respectively. 
\end{example}

\indent It is important to note that the design in \eqref{integer_code_design} is linear, and hence, ICs are sphere-decodable \cite{HaV2}. In particular, the $2 \times 2$ ICs have been shown to exhibit reduced sphere decoding complexity property \cite{HaV2}. In general, we observe that the advantages in the complexity reduction diminishes with larger dimensions. 

\begin{remark}
Note that the proposed design resembles those of the algebraic STBCs \cite{Perfect1}, \cite{Perfect2}. However, in this case, such an algebraic structure fails over the finite ring $\mathbb{Z}_{2^{m/2}}$ for $m \geq 1$.
\end{remark}

\subsection{Reduced Number of Processor Bits for ICs}
\label{sec3_subsec1}

Since the coefficients of the design and the entries of $\mathcal{S}$ belong to a finite subset of $\mathbb{Z}[i]$, the fixed-point encoding operations can be performed with reduced number of bits. Further, as $\mathcal{S}$ is a square $2^{m}$-QAM and $\alpha  = 2^{\frac{m}{2}}$, the in-phase and quadrature components of the complex numbers of the matrices are in the interval $[-d, d]$, where $d = 2^{\frac{mn}{2}} - 1.$ Hence, the proposed ICs can be encoded using $\frac{mn}{2} + 1$ bits per dimension with $2^{m}$-QAM constellation. For the special case of $2 \times 2$ MIMO, the proposed ICs can be encoded using $3$, $5$, and $7$ bits for $4$, $16$, and $64$-QAM constellations, respectively. In Table \ref{table_min_bits}, these numbers are compared with that of the Golden code and the Silver code \cite{HaV1}. Note that although we need $m/2$ bits (along each dimension) to represent the symbols of $2^{m}$-QAM, the numbers shown in Table \ref{table_min_bits} are the total number of bits required to carry out the linear operations in the design. For larger MIMO channels (i.e., with large values of $n$), the dynamic range for the complex numbers increases, which in-turn requires higher number of processor bits for the encoder. Thus, ICs provide significant reduction in the number of processor bits in small MIMO channels.
\begin{center}
\begin{table}
\caption{Minimum number of encoding bits for the Golden code, the Silver code, and the Integer code}
\begin{center}
\begin{tabular}{|c|c|c|c|c|c|c|c|c|c|c|}
\hline Constellations & Golden code & Silver code & Integer code\\
\hline 4-QAM & 7 & 6 & 3\\
\hline 16-QAM & 8 & 9 & 5\\
\hline 64-QAM & 9 & 9 & 7\\
\hline
\end{tabular}
\end{center}
\label{table_min_bits}
\end{table}
\end{center}
\subsection{Low PAPR of ICs}
\label{sec3_subsec2}
We use the following definition of the PAPR \cite{boyd} given by $\eta = \max |\textbf{X}_{j, k}|^{2}/E[|\textbf{X}_{j, k}|^2],$ where $\textbf{X}_{j, k}$ denotes the complex number transmitted through the $j$-th antenna at the $k$-th channel use. For ICs since $\alpha = 2^{\frac{m}{2}}$ and $\gamma = \imath$, the complex symbols transmitted through each antenna take values from a regular $M^{n}$-QAM constellation. As a result, the PAPR of the ICs is same as the PAPR of $M^{n}$-QAM. For perfect codes \cite{Perfect1}, it is well known that the transmitted complex numbers take values from a \emph{rotated} and \emph{unequally spaced} $2$-dimensional constellation, which in-turn results in larger value of PAPR when compared with that of the regular $M^{n}$-QAM. In Table \ref{PAPR_table}, we list the PAPR values of several full-rate codes when the information symbols come from $4$, $16$, and $64$-QAM constellations. For the values corresponding to $n = 3$, HEX constellation is used, while for $n = 2$ and $4$, regular QAM is used. The numbers in Table \ref{PAPR_table} highlight the PAPR advantages of the ICs especially for larger MIMO and larger QAM constellations.
\begin{center}
\begin{table}
\caption{PAPR (in decibel scale) for Integer codes and Perfect codes (GC- Golden code, PC- perfect code).}
\begin{center}
\begin{tabular}{|c|c|c|c|c|c|c|c|c|c|c|}
\hline MIMO & STBCs & 4-point & 16-point & 64-point\\
dimension &  & constellation & constellation & constellation\\
\hline n = 2 & GC & 2.77 & 5.32 & 6.45\\
\hline & IC & 2.55 & 4.21 & 4.62\\
\hline n = 3 & PC & 4.31 & 6.86 & 7.99\\
\hline & IC & 3.67 & 4.63 & 4.75\\
\hline n = 4 & PC & 5.74 & 8.29 & 9.42\\
\hline & IC & 4.22 & 4.73 & 5.59\\
\hline
\end{tabular}\begin{scriptsize}
•
\end{scriptsize}
\end{center}
\label{PAPR_table}
\end{table}
\end{center}
\subsection{Trace and Determinant Properties of ICs}
\label{sec3_subsec4}
\begin{figure}[!]
\centering
\includegraphics[width=3.1in]{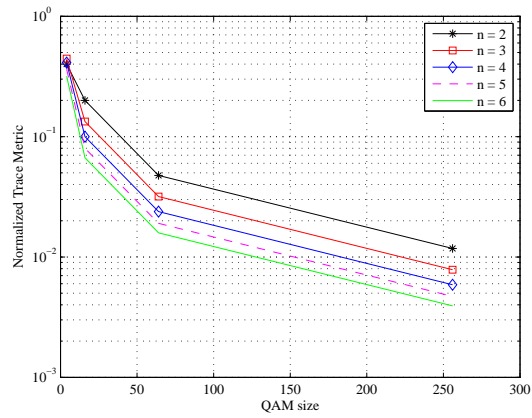}
\vspace{-0.5cm}
\caption{Normalized trace metric for Integer codes for several transmit antennas and QAM sizes (4, 16, 64, and 256-QAM).}
\label{ref_trace}
\end{figure}

\indent In the literature on STBC design, the minimum determinant criterion \cite{TSC} is used for high SNR error performance, whereas the minimum trace criterion \cite{YVXC}, \cite{TiH} is used for the low and moderate SNR error performance. For the ICs, we have adhered to the trace criterion by using the circulant matrix $\Phi$, which spreads the weight of $\alpha$ over different information symbols in every layer. For the specific case of $2 \times 2$ MIMO, the position of $\alpha$ is swapped in the design as $x_{1,1} + \alpha x_{1,2}$ and $\alpha x_{1,1} +  x_{1,2}$. A similar technique is adopted on the off-diagonal elements for $x_{2,1}$ and $x_{2,2}$. As shown in Table \ref{code_properties_table}, this technique results in a high value of the minimum trace metric. For larger MIMO channels, Fig. \ref{ref_trace} shows the minimum trace values which are normalized by the average transmit power (per channel use) of ICs. We point out that the ICs do not satisfy the minimum determinant criterion. However, the number of codeword pairs that do not satisfy the determinant criterion is very small. For the $2 \times 2$ MIMO, the percentage of such pairs is reported in Table \ref{code_properties_table} for several QAM constellations. We note that ICs do not (and are not designed to) achieve the well known diversity multiplexing trade-off property. Instead, ICs are practical codes that provide good results at low and moderate SNR values due to ({\em i}) a high value of the minimum trace metric, and ({\em ii}) a very low percentage of the zero determinant codeword pairs.

\begin{small}
\begin{center}
\begin{table*}
\caption{Determinant and trace properties of ICs for $2 \times 2$ MIMO}
\begin{center}
\begin{tabular}{|c|c|c|c|c|c|c|c|c|c|c|}
\hline Constellations & Minimum non-zero Det & Minimum Trace & $\#$ Distinct & Percentage of\\
& & & difference matrices & zero determinant matrices\\
\hline $4$-QAM &  $0.04$ & $0.4$ & 6561 & $0.487$ \%\\
\hline $16$-QAM & $0.1384 \times 10^{-3}$ & $0.2$ & 5764801 & $0.0011$ \%\\
\hline $64$-QAM & $5.3 \times 10^{-7}$ & $0.047$ & $2.5629 \times 10^{9}$ & $1.8729 \times 10^{-5}$ \%\\
\hline
\end{tabular}
\end{center}
\label{code_properties_table}
\end{table*}
\end{center}
\end{small}

\subsection{Simulation Results on the Error Performance of ICs}
\label{sec4}




\begin{figure}[!]
\centering
\includegraphics[width=3.1in]{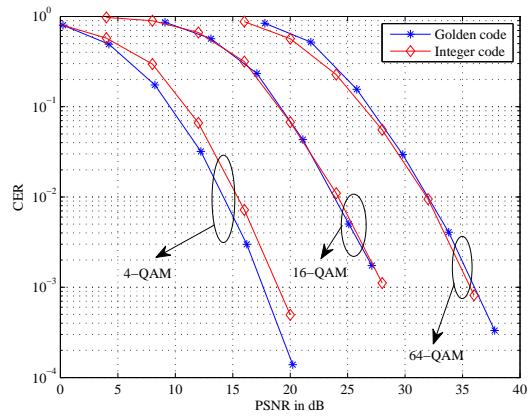}
\vspace{-0.5cm}
\caption{CER comparison of unquantized Golden code and Integer code for $2 \times 2$ MIMO with various QAM constellations.}
\label{ref_CER_2_X_2}
\end{figure}

\begin{figure}[!]
\centering
\includegraphics[width=3.1in]{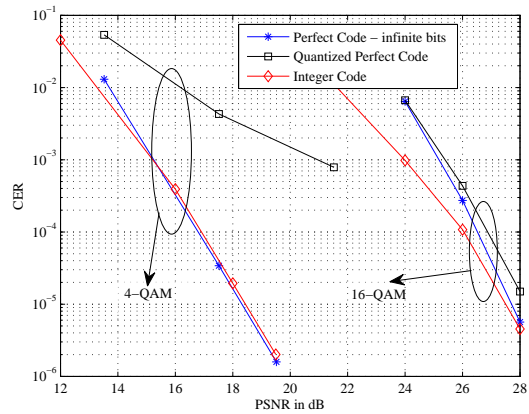}
\vspace{-0.5cm}
\caption{CER comparison of quantized Perfect code, unquantized Perfect code, and the Integer code for $4 \times 4$ MIMO with 4-QAM and 16-QAM. For quantized Perfect code and the Integer code, 5 bits and 9 bits are used for the encoder with 4-QAM and 16-QAM, respectively.}
\label{ref_CER_4_X_4}
\end{figure}


\indent In Fig. \ref{ref_CER_2_X_2}, we compare the codeword error rate (CER) of $2 \times 2$ ICs with the infinite-precision versions of the Golden code. We use PSNR (as defined in Section \ref{sec2}) to capture the PAPR differences among the codes under comparison. If we compare the ICs with the finite-precision versions of the Golden codes (with 3, 5, and 7 encoder bits for 4, 16, and 64-QAM, respectively), then ICs will outperform the Golden codes since the latter codes are known to experience error-floors at high SNR values \cite{HaV1}. Along the similar lines, in Fig. \ref{ref_CER_4_X_4}, we compare the CER of $4 \times 4$ ICs with the infinite-precision and finite-precision versions of the Perfect codes with 4 and 16-QAM. For the $4 \times 4$ ICs, we observe that the quantization advantage is prominent with $4$-QAM, while the PAPR advantage is prominent for 16-QAM. In conclusion, ICs, in addition to requiring fewer bits for the encoder, keeps the error performance quite close to that of the infinite-precision version of the perfect codes in the \emph{practical} SNR range of interest.

\section{Summary and Directions for Future Work}
\label{sec6}
We have proposed full-rate integer STBCs with reduced number of bits at the encoder and low PAPR values. We have shown that the reduction in the number of processor bits is significant for small MIMO channels, while the reduction in the PAPR is significant for large MIMO channels. An interesting direction for future work is to obtain an algebraic construction of full-rate integer STBCs over the ring $\mathbb{Z}_{2^{m/2}}$ for $m \geq 1$. Another direction for future work is to design ICs that are robust to quantization effects at the receiver.

\end{document}